\documentstyle[twocolumn,floats,epsf,aps,prl]{revtex}

\begin{document}

\twocolumn[\hsize\textwidth\columnwidth\hsize\csname 
@twocolumnfalse\endcsname

\title {\bf Substrate-adsorbate coupling in CO-adsorbed copper}
\author{Steven P. Lewis and Andrew M. Rappe}
\address{ Department of Chemistry and Laboratory for Research on the Structure 
of Matter, \\ University of Pennsylvania, Philadelphia, PA 19104.}
\date{\today}
\maketitle

\begin{abstract} 
The vibrational properties of carbon monoxide adsorbed to the copper (100)
surface are explored within density functional theory.  Atoms of the
substrate and adsorbate are treated on an equal footing in order to examine
the effect of substrate--adsorbate coupling.  This coupling is found to have
a significant effect on the vibrational modes, particularly the in-plane
frustrated translation, which mixes strongly with substrate phonons and
broadens into a resonance.  The predicted lifetime due to this harmonic decay
mechanism is in excellent quantitative agreement with experiment.
\end{abstract} 


]

An important consequence of molecular adsorption to a metal surface is the 
emergence of new, low-frequency vibrations associated with fluctuations of the 
chemisorption bond.  These ``external modes'' correspond to translations
and rotations of the free molecule which become frustrated upon 
adsorption to a substrate.  They play an instrumental role 
in many fundamental processes at surfaces, including chemical reactivity, 
desorption, and surface diffusion \cite{Somorjai94,Vanselow,SS30th},  and
are, therefore, of intense scientific interest.

In this letter, we theoretically investigate the vibrational dynamics of in-plane
frustrated translational (FT) motion for a prototypical adsorbed system: carbon 
monoxide on the (100) surface of copper at half-monolayer coverage.  Frustrated
translational vibrations are considered particularly 
important in surface chemistry because they are typically very low in energy (a few 
meV) and are thus thermally activated.  This mode has been found to be extremely 
short-lived for CO on copper, with a lifetime in the few picosecond range
\cite{Germer93,Culver93,Graham95}.  The mechanisms governing the relaxation of 
this mode are of considerable interest.

The present study focuses on the role of the substrate lattice in FT decay.   
Vibrational states for the combined substrate-adsorbate system are computed from
first principles, with all atoms treated on an equal footing.  These 
investigations reveal that the dominant contribution to FT relaxation comes from
resonant mixing of frustrated translations with long-wavelength bulk copper 
phonons.  This is a purely harmonic mechanism, and it gives rise to a computed 
lifetime in good quantitative agreement with experiment \cite{Germer93}.  To 
our knowledge, this is the first detailed, quantum-mechanical investigation 
for an adsorbed metal system that demonstrates the strong interaction between 
adsorbate motion and long-wavelength bulk phonons.  An earlier classical model 
of FT damping \cite{Persson85} considered this type of coupling for the case of 
an isolated FT oscillator attached to a semi-infinite, elastic medium.  The 
damping was expressed as a macroscopic frictional force.  Other microscopic 
theoretical studies have primarily considered only anharmonic effects in 
addressing adsorbate-lattice vibrational coupling, and have failed to identify 
this resonance because they have not modeled a large enough portion of the 
substrate to accommodate the long-wavelength phonons involved in resonant 
harmonic mixing.

Carbon monoxide on the (100) surface of copper has been the subject of a
recent pump--probe laser experiment \cite{Germer93} to investigate 
the role that FT motion plays in energy transfer between the substrate and the 
adsorbate overlayer.  This study measures the time-resolved vibrational 
response of the CO molecules to picosecond heating of the copper substrate, 
and interprets the resulting transient-response data as revealing the 
vibrational dynamics of FT motion.  This experiment observes a rapid vibrational
excitation followed by a slower decay with a characteristic damping time of 
$2.3\pm 0.4$~ps.  The authors model the vibrational relaxation by a two 
heat-bath model coupling adsorbate motion to electrons and phonons of the 
substrate.  The fitted relaxation times due to electronic and 
lattice coupling are $5.1\pm 0.4$~ps and $4.2\pm 0.7$~ps, respectively.
These measurements were taken at a substrate temperature of about 100~K and 
at a surface coverage of half of a monolayer.

%
\begin{table*}[t]
\caption{Computed equilibrium layer separations for (a) CO adsorbed 
to the copper $(100)$ surface, (b) the bare copper $(100)$ surface, 
and (c) a free CO molecule.  The comparison to bulk copper
refers to the computed separation between $(100)$ layers in
bulk copper: 1.796 \AA.
}

\begin{tabular}{ldddc}
Layer & \multicolumn{2}{c}{Separation (\AA)} & \% difference from & Experiment \\
Pairs & Covered Site & Empty Site            & bulk Cu spacing    &            \\
\hline
&&& \\
\multicolumn{5}{c}{(a) CO on copper $(100)$} \\
&&& \\
O--C           & 1.138 &       &      & 1.15$\pm$0.10 \AA\ \tablenote[1]{Reference \protect\cite{Ander79}} \\
C--Cu$_1$      & 1.852 &       &      & 1.92$\pm$0.05 \AA\ \tablenotemark[1] \\
Cu$_1$--Cu$_2$ & 1.805 & 1.768 &       & \\
Cu$_2$--Cu$_3$ & 1.804 & 1.821 &       & \\
Cu$_3$--Cu$_4$ & 1.809 & 1.792 &       & \\
&&& \\
\multicolumn{5}{c}{(b) Bare copper $(100)$} \\
&&& \\
Cu$_1$--Cu$_2$ & 1.771 &       & -1.38 & -(1.2--2.4)\% \tablenote[2]{Reference \protect\cite{Rodach93}} \\
Cu$_2$--Cu$_3$ & 1.807 &       & +0.60 & +(0.0--1.0)\% \tablenotemark[2] \\
Cu$_3$--Cu$_4$ & 1.795 &       & -0.02 & \\
&&& \\
\multicolumn{5}{c}{(c) Free CO molecule} \\
&&& \\
O--C           & 1.123 &       &       & 1.128 \AA \\
\end{tabular}

\label{struc}
\end{table*}
%

Vibrational damping in this system has also been addressed theoretically.  A 
recent quantum chemistry analysis \cite{Head92} has examined relaxation of the
adsorbate vibrational motions through nonadiabatic coupling to conduction 
electrons of the substrate.  In this study, the copper substrate is modeled 
as a cluster with atoms fixed in their experimental positions.  The coupling 
rates to the copper conduction electrons are determined using the Fermi golden 
rule, and the FT mode is found to decay slowly via this mechanism 
($\tau = 108$~ps) compared to other modes, such as the frustrated rotation 
(FR) and the internal CO stretch vibration ($\tau = 2-3$ ps).  The same 
authors have also performed a classical molecular-dynamics simulation of this 
system to include both vibrational and electronic damping mechanisms 
\cite{Tully93}.  Nonadiabatic electronic coupling is included via an electronic 
friction formalism, and the vibrational motion is described by empirical Morse 
potentials.  The substrate is modeled as a three-layer copper slab.  These 
calculations again predict a comparatively slow FT decay rate ($\tau = 14$~ps) 
at low temperatures.  As temperature increases above about 150 K, however, the 
computed decay time does enter the few picosecond range.

The vibrational properties of molecules interacting with surfaces have 
generally been described in terms of localized vibrational modes involving
only motion of the molecule relative to the surface.  Atoms of the substrate
are typically frozen in place, and the vibrational modes, by construction,
only span the atomic degrees of freedom of the adsorbate.  In this theoretical
study, however, we are explicitly interested in exploring the nature of the 
coupling between molecular and substrate vibrations, and thus we require an
unbiased theory which incorporates atomic motions of both components equally.
This is accomplished by computing interatomic harmonic force constants for the 
combined system, and solving the resulting coupled-oscillator equations to get 
the normal modes of vibration.

For the purposes of computing force constants, the adsorbed surface is modeled
as a seven-layer copper (100) slab with half-monolayer CO coverage of the slab 
surfaces in a $(\sqrt{2}\!\times\!\sqrt{2})R45^\circ$ pattern.  Full periodicity 
is imposed by periodically repeating the slab perpendicular to the surface with 
12 \AA\ of vacuum separating adjacent slabs.  Force constants are obtained by 
making a series of perturbations of the system away from its equilibrium 
configuration, and computing the resulting forces on all of the atoms, from first 
principles.  Long-wavelength bulk copper phonons are included in the calculation
by defining a much thicker effective slab \cite{Chen91}, in which 1000 bulk-like 
layers are inserted in the middle of the seven-layer slab.  Interlayer force 
constants for bulk copper are computed once in a separate calculation, and thus 
little additional computational effort is required to set up the coupled-oscillator 
equations for a much more realistically sized system.

The first-principles calculations are performed within density functional theory 
\cite{Payne92}, and the local density approximation (LDA) is employed to describe 
electron-electron interactions.  Nonlocal corrections to the ex\-change-correlation 
functional have been shown not to have a significant influence on structural and 
vibrational properties in this system \cite{Philipsen94}.  Optimized 
pseudopotentials \cite{Rappe90} are used to model electron-ion interactions.  
The one-electron wave functions are expanded in a well-converged plane-wave basis 
containing all plane waves up to a 50 Ry cutoff energy.  Integrations over the 
surface Brillouin zone are performed on a discrete mesh of 8 k-points with 0.3 eV 
Gaussian broadening of energy levels. 

%
\begin{figure*}[t]
\epsfysize=2.75in
\centerline{\epsfbox[5 430 592 675]{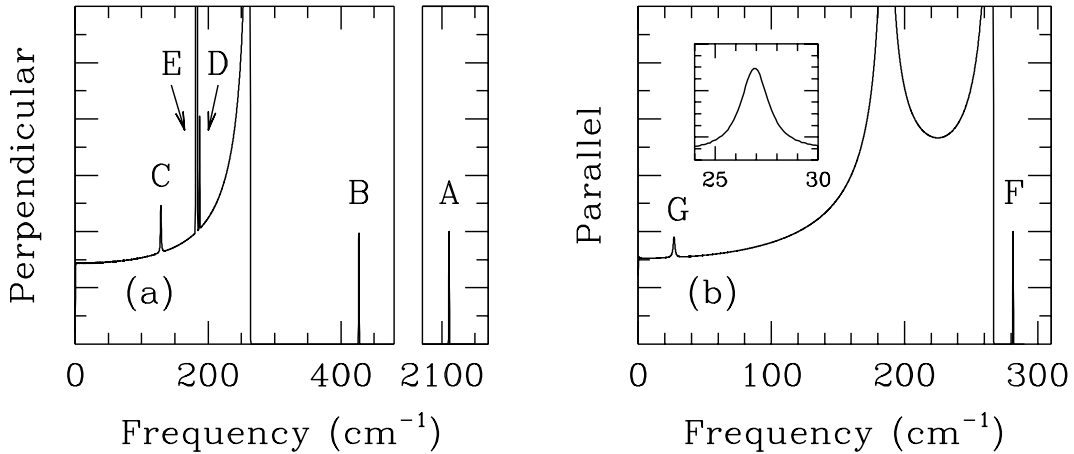}}
\caption{
Density of (a) perpendicular and (b) parallel vibrational modes for a
$(\protect\sqrt{2}\!\times\!\protect\sqrt{2})R45^\circ$ overlayer of CO 
on the copper (100) surface.  The inset in (b) highlights the frustrated 
translational resonance centered at 27 cm$^{-1}$.  The modes correspond to 
the $\bar{\Gamma}$-point in the surface Brillouin zone.  Arbitrary units 
are used for the density of modes.
}
\label{vib}
\end{figure*}
%

An analysis of the vibrational states of CO on copper requires first 
determining the equilibrium atomic positions.  These calculations are summarized 
in Table \ref{struc}.  The O-C and C-Cu bond lengths for the adsorbed molecule 
are in excellent agreement with experiment \cite{Ander79}.  For comparison, the 
structures of a bare copper (100) surface and a free CO molecule are also 
presented.  At an adsorption site, the copper atom relaxes outward relative 
to the clean surface by about 1.9\%, whereas the empty-site coppers remain 
essentially fixed.  Furthermore, the bond length of CO expands relative to 
the gas-phase value by about 1.3\%.

Vibrational modes at the $\bar{\Gamma}$ point of the 
$(\sqrt{2}\!\times\!\sqrt{2})R45^\circ$ surface Brillouin zone (SBZ) have been 
computed and are illustrated in Fig.\ \ref{vib}.  The spectra are rich in detail, 
and are in good quantitative agreement with the available experimental data.  
This serves as strong evidence of the validity of local density 
functional theory for describing the vibrational properties of this system 
accurately.  Symmetry allows the modes to be classified as having atomic motions 
either perpendicular or parallel to the surface.  The identity of the modes is 
established by examining the normal-mode polarization vectors.  The density of 
perpendicular modes is shown in Fig.\ \ref{vib}(a).  It is characterized by a 
broad band of modes below 263 cm$^{-1}\,$ and sharp features labeled $A$--$E$.  The two 
isolated modes labeled $A$ and $B$ are strongly localized at the adsorbate and 
constitute the O-C and the C-Cu bond-stretching modes, respectively.  The computed 
frequencies, 2111 cm$^{-1}\,$ and 427 cm$^{-1}\,$, respectively, compare favorably with the 
experimental values of 2085 cm$^{-1}\,$ \cite{Ryberg82} and 345 cm$^{-1}\,$ \cite{Hirsch90}.  

The features labeled $C$ and $D$ correspond to the surface phonons, $S_1$ and
$S_2$, respectively, at the $\bar{M}$ point of the underlying bare copper SBZ
\cite{Chen91}.  This point folds back to the zone center in the adsorbed overlayer 
structure.  The $S_1$ feature is centered at 129 cm$^{-1}\,$, which is in excellent
agreement with a helium-atom scattering (HAS) measurement of 123 cm$^{-1}\,$ for this system 
\cite{Ellis95}.  The $S_2$ mode has a frequency of 187 cm$^{-1}\,$, which can be compared 
to the measured bare-surface value of 163 cm$^{-1}\,$ \cite{Chen91}.  To our knowledge,
no experimental results for this mode have been reported for the adsorbed surface.
Finally, the broad band below 263 cm$^{-1}\,$ and feature $E$ correspond to bulk
copper phonon modes that project onto the $\bar{\Gamma}$ and $\bar{M}$ points,
respectively, of the bare SBZ.  The latter band is known to be very narrow
\cite{Nicklow67}, resulting in a very large peak in the density of modes.

The density of parallel modes is shown in Fig.\ \ref{vib}(b).  It consists
of a broad band below 266 cm$^{-1}\,$ and two features labeled $F$ and $G$.  The
broad band corresponds to bulk copper phonons, with modes below (above) 
185 cm$^{-1}\,$ coming from the $\bar{\Gamma}$ ($\bar{M}$) point of the bare SBZ.
The isolated mode labeled $F$ is highly localized on the adsorbate and
corresponds to FR motion of the CO molecule.  Its computed frequency is 
282 cm$^{-1}\,$, in excellent agreement with the 285 cm$^{-1}\,$ experimental value 
\cite{Hirsch90}.

\begin{figure}[ht]
\epsfysize=2.44in
\centerline{\epsfbox[5 412 592 630]{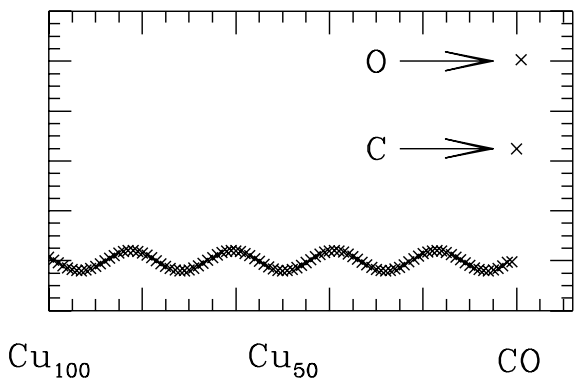}}
\caption{
Illustration of a typical mode in the frustrated translational resonance peak.
The vertical axis gives the relative in-plane atomic displacements for the mode, 
and the horizontal axis labels the atomic layers.  Motion of the adsorbate atoms
and 100 layers of the copper substrate are shown.
}
\label{ft}
\end{figure}

Feature $G$, shown in more detail in the inset of Fig.\ \ref{vib}(b), is associated 
with FT motion of the CO molecule.  It contains modes exhibiting the large-amplitude, 
in-phase wagging motion of carbon and oxygen atoms that is characteristic of frustrated 
translations.  However, an appreciable amplitude is found to persist throughout the 
substrate, with the copper atoms undergoing bulk phonon motion.  The atomic displacements 
for a typical normal mode in this peak are illustrated in Fig.\ \ref{ft}.  This behavior 
demonstrates that FT motion is not a normal mode of CO on copper, but rather it mixes 
with bulk copper phonons to form a resonance.  This resonance has a spectral width of 
1.8 cm$^{-1}\,$ and is centered at 27 cm$^{-1}\,$.  HAS measurements find a 
$\bar{\Gamma}$-point FT frequency of 32 cm$^{-1}\,$ for this system \cite{Ellis95}, 
indicating excellent agreement between theory and experiment.

%
%

The resonant broadening of the FT mode through coupling to bulk phonons provides
a purely harmonic description for the observed FT relaxation.  In this picture, 
initial heating of the substrate excites a vibrational wave packet that is 
localized near the surface.  Its normal mode components evolve in time according
to different frequencies, causing the wave packet to propagate and broaden over 
time.  In this way, vibrational energy initially deposited near the surface 
spreads throughout the substrate.  The spectral width of the FT resonance
peak in Fig.\ \ref{vib}(b) gives a measure of the decay rate of the mode and
implies a lifetime of 3.0 ps.  This excellent quantitative agreement with the 
experimental decay time of $2.3\pm 0.4$ ps at T$\approx$100 K is strong evidence 
that resonant mixing is the dominant mechanism of FT relaxation in this system.  
This conclusion is consistent with the theoretical work of Ref.\ \cite{Head92}, 
which finds that FT decay via nonadiabatic electronic coupling is quite slow
for this system ($\tau = 108$~ps). 

In this density-functional study, we have explored the vibrational coupling
between a CO adsorbate overlayer and a copper substrate lattice.  By treating 
all atoms of the combined system on an equal footing and including enough bulk 
copper to describe long-wavelength substrate phonons, we have shown that the 
vibrational properties of this system are significantly affected by 
adsorbate--substrate interactions.  In particular, we find that the FT mode mixes 
with long-wavelength, transverse phonons of the substrate to form a resonance.  
This resonant mixing provides a microscopic description of FT relaxation involving 
propagation of a wave-packet of harmonic modes.  Based on this mechanism, we 
predict a decay time of 3.0 ps, in excellent quantitative agreement with 
experiment.  We thus conclude that harmonic resonant mixing with substrate phonons 
is the dominant mechanism of FT damping in this system.  These calculations 
highlight the need for a sophisticated treatment of the substrate in achieving 
a quantitative understanding of the properties of adsorbed metal surfaces.

The authors would like to acknowledge valuable discussions with E. J. Mele, 
H.-L. Dai, J. P. Culver, and R. M. Hochstrasser.  Computational support for 
this project was provided by the San Diego Supercomputer Center.

\end{document}